# Photoelastic Stress Response of Complex 3D-Printed Particle Shapes

[1*]Negin Amini, [1]Josh Tuohey, [1]John M. Long, [1]Jun Zhang, [1]David A. V. Morton [2]Karen Daniels, [2]Farnaz Fazelpour and [1]Karen P. Hapgood

[1]School of Engineering, Deakin University, 75 Pigdons Road Waurn Ponds, Geelong, VIC 3216, Australia

[2]Department of Physics, North Carolina State University, Raleigh, North Carolina, USA.

[*]Corresponding author: negin.a@deakin.edu.au

Abstract

While stress visualization within 3-dimensional particles would greatly advance our understanding of the behaviors of complex particles, traditional photoelastic methods suffer from a lack of available technology for producing suitable complex particles. Recently, 3D-printing has created new possibilities for enhancing the scope of stress analysis within physically representative granules. Here, we investigate and evaluate opportunities offered by 3D-printing a single particle with a complex external shape with photoelastic properties. We report the results of X-ray computed tomography and 3D-printing, combined with traditional photoelastic analysis, to visualize strain for particles ranging from simple 2D discs to complex 3D printed coffee beans, including with internal voids. We find that the relative orientation of the print layers and the loading force affects the optical response of the discs, but without a significant difference in their mechanical properties. Furthermore, we present semi-quantitative measurements of stresses within 3D-printed complex particles. The paper outlines the potential limitations and areas of future interest for stress visualization of 3-dimensional particles.

1       Introduction

Photoelastic methods have been widely used in experiments to identify regions of stress within single and bulk 2D particle systems [1]. A sample photoelastic visualization is shown in Figure 1: the stress applied on a polymer disc results in an alternating bright/dark fringe pattern when viewed using polarized light. The earliest report of this method being used in the study of granular materials was in the late 1930s, where it was applied to powdered glass in a Christensen filter [2]. In the 1950s, a report by Wakabayashi applied photoelastic methods to determine stress in a powder mass [3]. Researchers have since drawn on the technique to further their understanding of the internal forces within granular materials, where localized force transmission occurs through structures known as *force chains* [4, 5]. The bulk behaviors of real-world particulate matter exhibit similar internal patterns of stress.

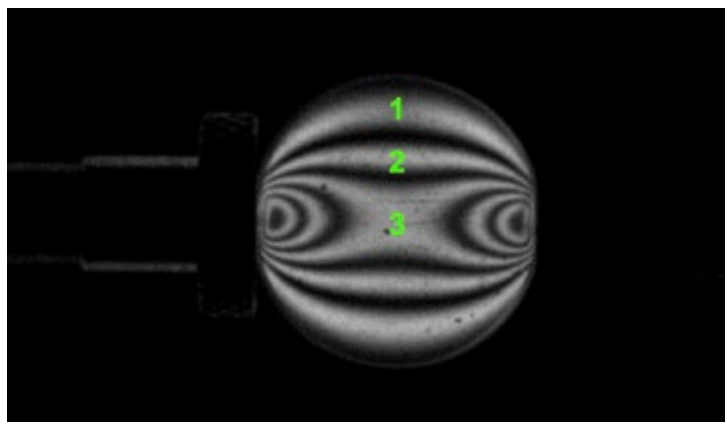

**Figure 1. Image illustrating the photoelastic response of a disc subject to diametric load under darkfield configuration with the isochromatic fringes numbered [6].**

To date, quantitative studies [6, 7] have largely focused on 2D particle systems, primarily flat discs [8], but also non-spherical shapes such as ellipses [9] and pentagons [10]. While there have been recent experiments on monolayers of photoelastic spheres [11-13], it remains a challenge to interpret the resulting fringe patterns, even when the system is index-matched. In response to these experimental limitations, there have been advancements in computational modelling using Finite Element Analysis (FEA). This method is capable of efficiently analyzing stress transmission on complex particle systems in a relatively short timeframe [14] and can predict the shapes and forces of deformable particles [15].

The underlying mechanism of the photoelastic technique is the double refraction of the polarized light incident on transparent polymer, which is sensitive to the local stress along the optical path [16]. The simplest example is the photoelastic response for diametric loading on disc shaped particles (see Figure 1), which has been well established analytically [1]. For this shape, the number of observed bright and dark fringes increases with the force applied to the particle and can be used as a means of calibration for quantitative analysis (Supplementary video 1). The intensity $I(x,y)$ at a particular point on the disc can be calculated by using equation 1 [17],

$$I(x,y) = I_o \sin^2 \frac{\pi(\sigma_1 - \sigma_2)}{F_\sigma} \qquad (1)$$

where ($\sigma_1 - \sigma_2$) is the principal stress difference at that point. Under monochromatic illumination by light of wavelength $\lambda$, the object-specific constant $F_\sigma = \lambda/Ch$ governs the response. Here, the thickness of the flat object is given by $h$, and the material's stress optic coefficient is $C$. It is important to note that when a disc is subjected to 2D stress, the fringe order is proportional to the thickness [1].

The gradient-squared ($G^2$) method [17] is a semi-quantitative measurement method for quantifying the 2D stress on a particle with an arbitrary number of vector contact forces, developed by Behringer *et al.* and collaborators. It is calculated by taking the local (pixel-wise) gradient, squaring, and averaging over all pixels in a particle or system, according to the equation (2) [18-21]:

$$G^2 = \sum_{i,j} \left[ \left(I_{i+1j} - I_{i-1j}\right)^2 + \left(I_{ij+1} - I_{ij-1}\right)^2 + \frac{1}{2}\left(I_{i+1j+1} - I_{i-1j-1}\right)^2 + \frac{1}{2}\left(I_{i+1j-1} - I_{i-1j+1}\right)^2 \right] \quad (2)$$

Although the $G^2$ method is not based directly on the underlying mechanism of photoelastic response, it has been widely used for quantifying stresses in granular systems because of its simplicity and ease of calculation [5, 20]. The $G^2$ method simply uses the image of the sample or particle being studied, and can be applied to particles of any shape, including non-circular particles [10], although with challenges in interpreting the results for angular shapes.

Until recently, most photoelastic stress studies used flat 2D particles because these (1) provide the simplest interpretation of the optical patterns and analytical solutions and (2) the 2D can be easily machined from flat sheets or cast in molds, although some post-processing treatment to remove residual stresses from the material is required. However, the recent introduction of 3D-printing using photoelastic materials has opened up new possibilities. *Lherminier et al.* [22] investigated the slow shearing of a compressed monolayer of 3D-printed discs made on the Objet 30 in Durus White 430 material. They successfully reproduced several features of real earthquakes through the dynamic nature of the force networks. Wang *et al.* [23] led a comprehensive characterization of the commercial 3D-printing material VeroClear, which has photoelastic properties, and is used in standard Stratasys Connex Objet printers [24] and used this material to 3D print rock-like structures based on x-ray tomography. The internal structure and mechanical response of natural reservoir rocks vary widely due to the sedimentary forming process, and understanding their behavior under stress is important in many geo-mechanical applications [23]. They reported that under triaxial conditions, the samples with horizontal layers have a higher stiffness compared with the vertical layers.

However, little variance was found in compression strength under different confining pressures. Importantly, this study was able to reproduce the ratio of the geological stress/rock modulus with a similar lab stress/polymer modulus. Similarly, Ju *et al.* investigated 3D-printed rock that contained inherent internal fractures for frozen stress and photoelastic tests. They also reported that the isotropic mechanical properties of the rock sample were close to that of the 3D-printed replica rock [25].

Breakage and deformation studies of real particles require either a very large number of samples to be tested, or an acceptance of large experimental variability and tolerance of uncertainty when interpreting the results. 3D-printing can be used to replicate real particle shapes naturally formed in the environment, as a tool for understanding the stress behavior of these complex particles and systems. In addition, 3D printing can create multiple identical copies of real irregular particles to be used in systematic experimental studies of stress and deformation, which overcomes a significant experimental hurdle. The option to create replica particles with photoelastic materials via 3D-printing permits even more sophisticated experiments to observe stresses for different particles under different stress conditions. Even a small change in the dimensions of the particle shape can have a significant impact on the stress and photoelastic responses. In the long term, the ability to visualize stress responses in 3D assemblies of complex 3D particles would facilitate better understanding of particle and agglomerate breakage in industrial applications. Although the development of enhanced models for 3-dimensional stress analysis of complex particle systems [25] is some way off, it is worth investigating what new information and analysis might be possible by 3D printing photo elastic materials into a more complex 3D particle shape.

In this study, we investigate the photoelastic response of a naturally-occurring complex particle -- a 3D coffee bean – including internal and external surfaces with positive and negative curvature. The presence of internal voids significantly complicates stress visualization, and therefore qualitative and quantitative tests using available methods were also conducted on a modified bean that contained no internal voids. A cross-section of the bean containing the voids was digitally extruded to observe the 2D photoelastic response. We also report the optical and mechanical properties of 3D-printed discs with different alignments of the print layer, compared to a control disc produced by traditional manufacturing methods as a comparative benchmark.

2      Material and Methods

2.1    Particle shape construction

The cylindrical disc shape (diameter 20 mm, thickness $h$ =10 mm) was created using Autodesk Inventor CAD software and exported as an *.STL file for 3D-printing. To reproduce the complex external and internal shape of a coffee bean, X-ray Computed Tomography (Zeiss Xradia 520 Versa XCT) was used to create a volumetric representation of a real coffee bean which contains internal voids. This technique records projection images of the x-ray energy transmitted as the sample is slowly rotated 360 degrees in 1-degree increments. The voltage and current required by the x-ray beam [26] is dependent on the sample thickness and material type; for this study, the optimum contrast of the coffee bean required 40 kV and 74 mA.

Materialise MIMICS software was used to reconstruct the XCT output DICOM images into a closed surface mesh *.STL file ready for 3D-printing [26]. The original size of the reconstructed coffee bean (11 x 4 x 7 mm) was doubled, and the rounded ends truncated to create a flat surface at the top and bottom of the coffee bean shape to assist with handling and applying loads during compression testing. Three versions of this coffee bean particle were investigated; one containing its internal void structure, another simplified version with internal voids removed, and a 2D extruded cross-section of a slice of the bean with internal voids.

2.2    Particle materials

The 3D-printed particles were made using Stratasys VeroClear UV-curable polymer, an acrylate-based photopolymer with photoelastic properties. For this study, the mechanical properties were experimentally measured using the ASTM D638 standard tensile and ASTM D695 standard compression tests and the data summarized in Table 1. The tensile specimen of Type-I design for the yield and tensile strengths and the standard compression cylinder specimen design for the elastic modulus were used [27, 28]. We investigated three orientations for the build of the tensile specimen and two orientations for the compression specimen on the 3D printer. Figure 2 shows the position of each specimen on the platform so that the print layers are parallel (Fig. 2A and 2D) or perpendicular (Fig. 2B, 2C and 2E) to the platform when they are when in the upright orientation. These tests measured the effect of print layer orientation on mechanical properties and provided material constants for use in the FEA analysis.

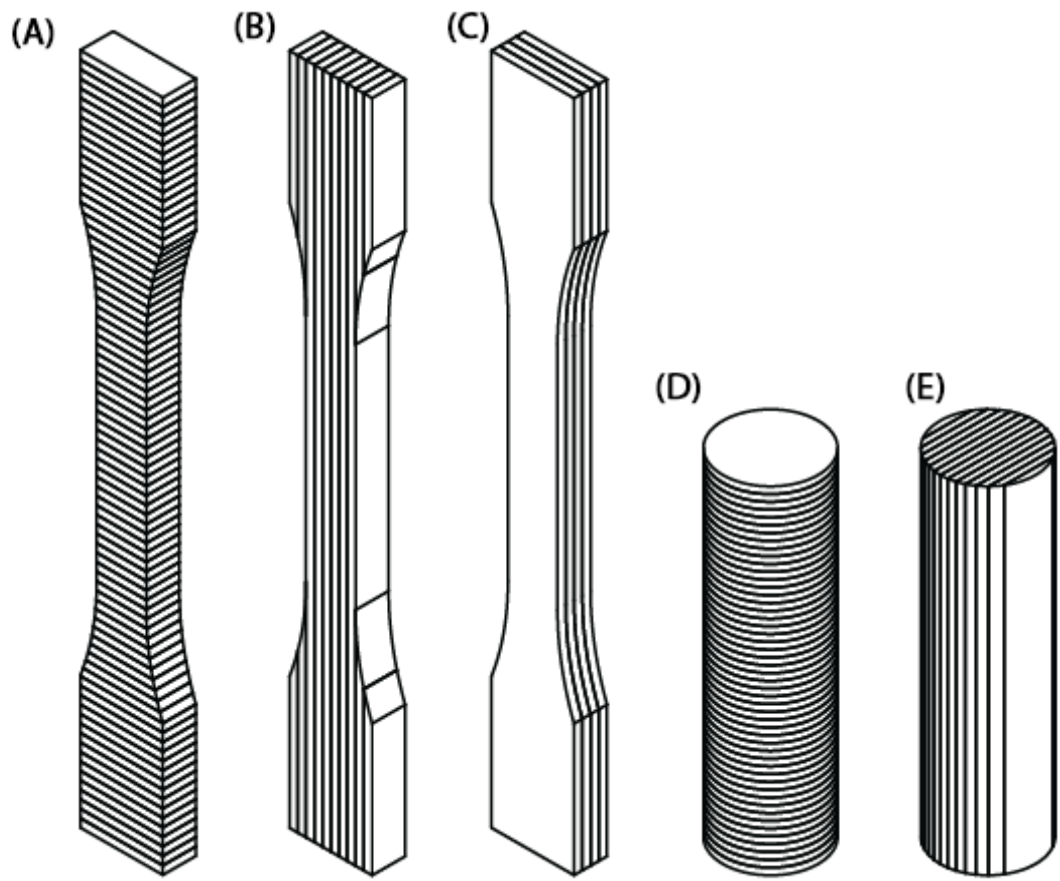

**Figure 2. Schematic illustrating the arrangement of the print layers for each specimen tested. (A) Tensile specimen - S1 (B) Tensile specimen - S2 (C) Tensile specimen - S3 (D) Compression specimen – S4 and (E) Compression specimen - S5.**

2.3  3D-printing of particles

A Stratasys Objet500 Connex3 3D printer was used to additively manufacture VeroClear particles using polymer jetting. This process involves selective jetting and polymerization of a photo-polymer resin to build up layers of a 3-dimensional object. Input geometry *.STL files were prepared for printing using Objet Studio slicing software to set build parameters and generate support structures. High quality 16μm layers at an XY-resolution of 600 DPI were specified. Compatible water-soluble support material was used to support overhanging regions and any internal voids; this material was removed in post-manufacture processing. In samples with internal voids, a small amount of support material was fully encapsulated within the internal voids and could not be removed. This was considered to have negligible impact on consequent test results as reported in [23, 25]. Sample particles were also printed without internal voids (completely solid) to eliminate this minor effect.

Flat discs were 3D-printed in the same manner as the ASTM compression standard specimens to determine whether the orientation of the print layers affects the optical properties of the particle. One disc typically takes 33 min to print in high-quality mode and requires six grams of material and two grams of support material. When multiple discs are printed simultaneously the machine requires less time: for example, 44 discs require two hours to print. Similarly, one coffee bean requires four grams of material and two grams of support and takes 35 minutes to print while 40 beans require one hour to print. An Olympus SZ61 DP22 macroscope was used to capture images of the resulting particles.

2.4    Laser cutting of disc particles

The control particles were laser cut from 10mm sheets of PMMA (polymethyl methacrylate) using a Trotec Speedy 500 laser cutter with 120-watt $CO_2$ laser. Adobe Illustrator was used to create the 2D profiles then exported as .DXF files for input to the laser cutter.

2.5    Instron compression of particles using the circularly polarized light configuration

For our polariscope, we use a brightfield configuration: two circularly polarizing filters (CIR-PL), both with the same chirality, are placed on either side of the sample (Fig. 3). Each CIR-PL filter comprises a quarter-wave plate on the side facing the particle, in series with a linear polarizer on the other side. An opaque acrylic sheet was placed between the monochromatic sodium lamp (wavelength 589 nm) and the first CIR-PL for uniform lighting.

Individual particles were placed between the steel plates on an Instron 5959 electromechanical universal test machine (50kN load cell). A strain rate of 1 mm/min was specified, up to a maximum displacement of 2.0 mm. Three specimens of each type were individually tested, while load and displacement were recorded at a rate of 10 Hz.

The images (TIFF, 2048x2048 pixels) were recorded with a PCO.edge 4.2 monochrome high-speed camera with a rolling shutter at 10 frames per second. A black acrylic cage with two holes for placement of the filters was placed around the compression plates for two reasons. First, as a safety measure to prevent flying shards in the instance of particle compression failure. Second, to restrict non-polarized light from interfering with the imaging process.

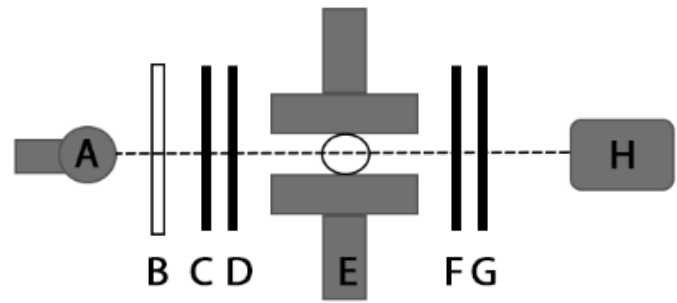

**Figure 3. Circular polarized light with a brightfield configuration set up around the Instron. A. Monochromatic sodium lamp, B. Light diffuser (acrylic sheet), C. Linear polarizer, D. ¼ wave plate, E. compression plates, F. ¼ wave plate, G. Linear polarizer, H. Camera recorder**

2.6     Gradient-squared ($G^2$) method for semi-quantification of 2D stress on system

The MATLAB Image Processing Toolbox was used to segment the TIFF images and isolate the foreground particle from the background and calculate the $G^2$ value (Eq. 2) on the visible stress fringes. The $G^2$ method was applied to the control disc and all 3D-printed particles in 2D and 3D. These values were then plotted against the corresponding measured test load, with the load normalized by the average cross-section area perpendicular to the load direction. For the disc, the load was divided by the average cross-sectional area perpendicular to the loading direction. This was done analytically to get an average area. For the coffee bean, (both cross-section and 3D variants) the load was divided by the average cross-sectional area of a finite number of intersecting planes.

2.7  Finite-Element Analysis of the particle compression

To visually validate the observed experimental stress fringes for the disc and the 2D extruded bean cross-section, Finite Element Analysis was applied. Quasi-static 2D plane stress FEA was conducted on the particles using ANSYS Mechanical. The VeroClear material was modeled as an isotropic, linearly elastic material with modulus 2500 MPa and Poisson ratio 0.38. Loads were applied using zero-slip contacts with flat rigid plates, the lower plate being fixed and the upper plate having a ramped 2 mm displacement. Solutions were obtained using the direct MAPDL solver and the Newton-Raphson method, which accounts for non-linearities associated with contacts and relatively large deformation.

We performed a mesh convergence study that monitored the reaction force at the displacement boundary conditions and found that a mesh-independent solution was found for uniformly-sized second-order quadrilateral elements (maximum nominal size 0.125 mm for the cylinder and 0.25 mm for the bean cross-section). To strike a balance between mesh-independence and solving time, we used 0.15 mm and 0.2 mm elements for the cylinder and digitally extruded bean cross-section, respectively. Additional refinement was added around regions with large stress gradients.

### 2.8 Eliminating the 3D-printed coffee bean shadows

Autodesk Maya software was used to construct a virtual representation of the polariscope test environment, using reference images to approximately match lighting, material, and camera setup. The Monte-Carlo based renderer Arnold was then used to render images of test specimens placed within a medium of refractive index of $n = 1$ (air) or $n = 1.52$ (index-matched to particle) on the initial unloaded appearance of the specimens. The rotation of polarization due to the waveplate-like behavior of acrylic was not accounted for.

## 3 Results and Discussion

### 3.1 Stress visualization of a 3D-printed disc

We first examined the photoelastic response of cylindrical discs to test for any optical effects arising from the layers created by the 3D-printing process. We performed the same measurements on a control disc, laser cut from a solid acrylic sheet. Images of the 3D-printed and laser cut discs before compression in natural light are shown in Figure 4A and 4B, respectively. We observe that the laser-cut disk is transparent, as expected, but the layers cause the 3D-printed discs to be translucent; both materials possess photoelastic properties.

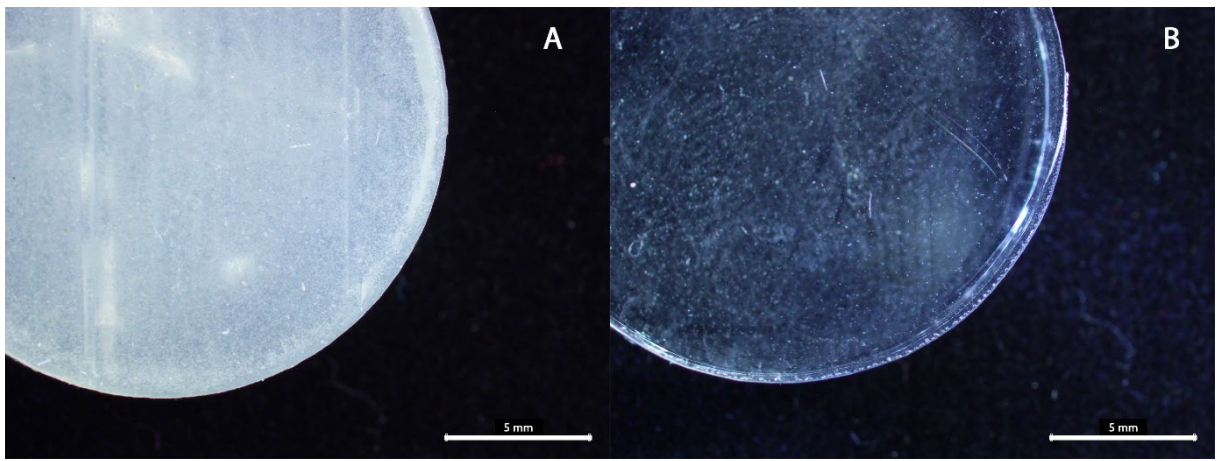

**Figure 4. Image of discs (dimensions: thickness 10 mm, diameter 20 mm) procured by two different methods (A) 3D-printed in VeroClear and (B) laser cut from acrylic sheets.**

One potential limitation of polymer 3D printing is the inherent anisotropy of 3D-printed particles due to the layer-based photo-polymerization building process. These layers could potentially lead to different optical and mechanical properties based on the print layer orientation and shape of the particle under compression. Furthermore, there may be effects due to the relative angle between the loading axis and the printing axis, such that a single set of material and optical parameters cannot be defined for the particles. Figure 5 shows the particle build positions on the 3D printing platform and a schematic of the particle layer orientations between the compression plates.

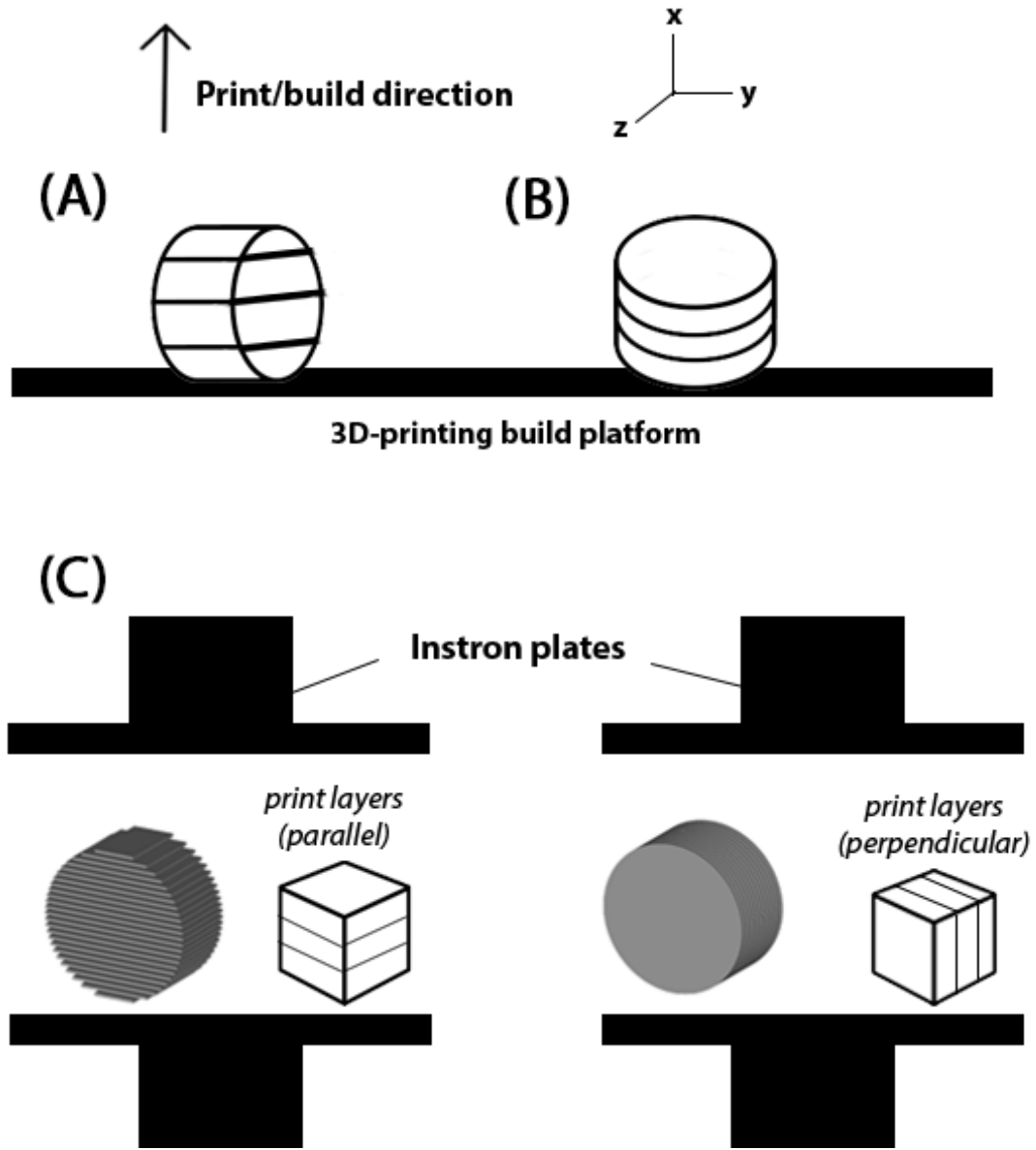

**Figure 5. A schematic of the particle orientations and the direction of build on the 3D printer platform showing (A) parallel and (B) perpendicular print layers when the discs are in the upright position. (C) the Instron setup showing the particle print layer arrangement with respect the compression plates.**

The orientation shown in Figure 5A comprises of 1250 layers stacked in the direction of the height of the disc (20mm), whereas Figure 5B contains half the number of layers stacked to make the thickness of the disc (10mm). When the particles are placed between the compression plates, they are viewed through a *brightfield* setup.

Figure 6A shows the image of an unloaded particle with parallel print layers, starting off *dark* prior to compression. In contrast, Figure 6B shows the image of an unloaded particle with perpendicular print layers which starts off light, as expected to be observed with a brightfield setup. Acrylic-based polymers (such as VeroClear) act as optical waveplates and rotate the polarization of light, unlike more favorable materials typically chosen for photoelastic materials. This rotation, in combination with the microstructure arising from the two different print layer orientations, makes the resulting discs differ in their optical response even when unloaded. If these particles were to be used for quantitative force measurements, it would be necessary to account for this baseline optical rotation using a calibrated waveplate, specific to the chosen print-orientation and particle thickness [1]. Finally, it is important to note that the stripes observed on Figure 6B arise from the traversing 3D printing nozzle dragging across the printing surface and are not the layers themselves.

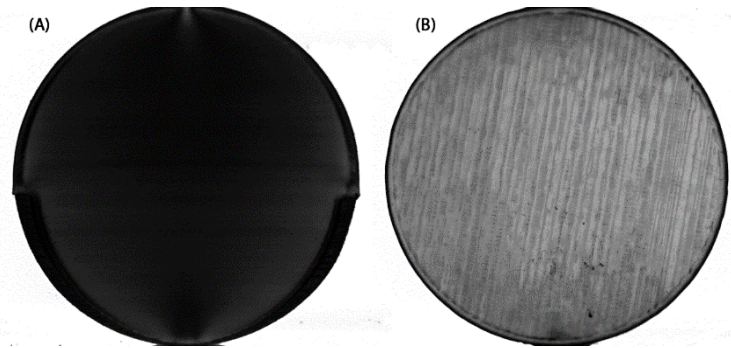

**Figure 6. Image of an unloaded disc viewed through the brightfield configuration with print layers (A) parallel and (B) perpendicular to the compression plates.**

To simplify the issue of anisotropy, only four discrete orientations were explored to examine whether there is an effect on the optical properties of 3D-printed discs during compression. The top row of Figure 7 illustrates these four orientations with the placement of all discs (3D-printed and control) as they are between the compression plates.

The middle row in Figure 7 shows the very first recorded isochromatic fringe ($N$=1) observed for each disc. The alignment of the first fringe with respect to the applied diametric contact force varies for each of the 3D-printed discs, due to the relative orientation of the layers and the loading axis. The 3D-printed disc with print layers perpendicular to the steel plates (Fig 7D) show a dark stress fringe which is comparable to the control disc (Fig 7E). The disc with parallel layers (Fig 7A) also shows the first recorded bright fringe to be similar; however, the outer edge of the disc contains an uncharacteristic silhouette partly due to the further rotation of polarization of light and the print layer orientation. Furthermore, Fig 7B and 7C show a slightly distorted/skewed first recorded bright fringes as well as an additional uncharacteristic silhouette on the outer edge. Based on these observations, we chose to print the more complex coffee bean shape so that all the layers were perpendicular to the steel plates, to minimize additional fringe distortion.

The bottom row in Figure 7 shows all discs with 5 fringes ($N$=5). The shape of the bright/dark bands appear to be the consistent with what has been reported in literature [6]. Force measurements were taken for $N$=5 fringes to check whether the shape of the particle has a significant effect on the material properties. A summary of the force load (*Newtons*) and displacement (*mm*) required to reach $N$=5 is given in Table 2. The control disc reached this point at almost three times higher force load than the 3D-printed discs, due to its stiffer material properties. The force required for the 3D-printed discs to reach $N=5$ was consistent regardless of print layer alignment. This is further discussed in Section 3.2 with the FEA data. For the discs with parallel print layers (Fig. 7A, B and C), the shape of the bright/dark bands on the outer edge of the discs is pushed further to the edge with each addition of more fringes. This is not consistent with the expected behavior for isotropic, elastic materials.

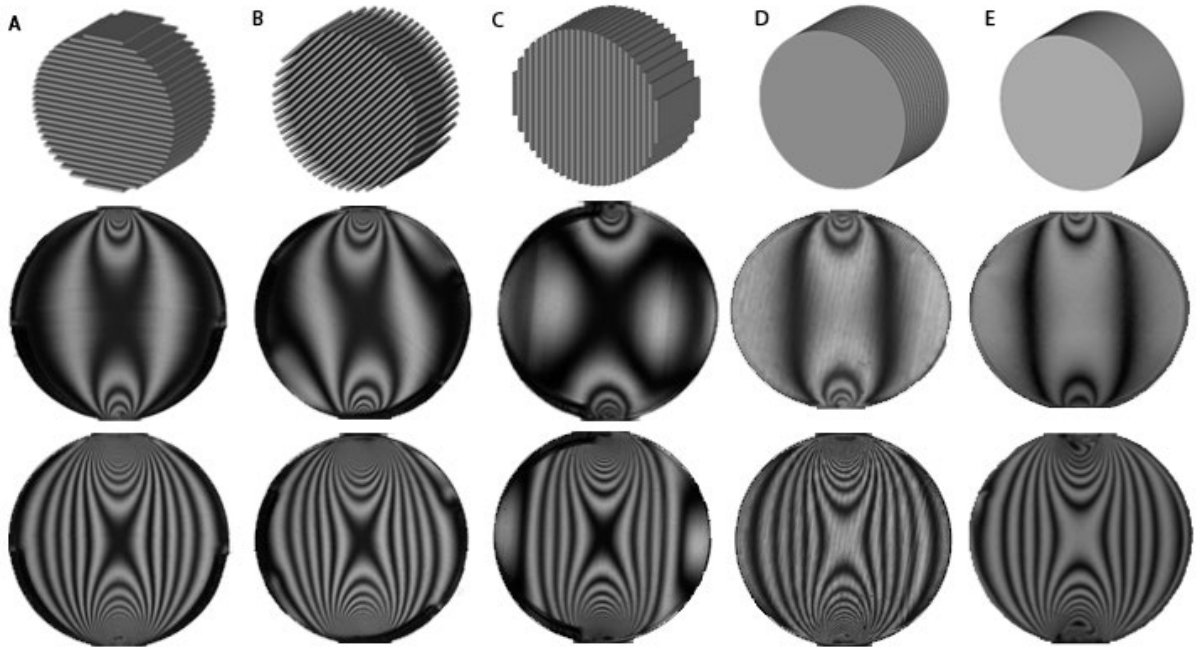

**Figure 7.** Top row – 3D representations of the discs to show print layer orientation with respect to the steel plates (A) parallel - 0° (B) parallel - 45° (C) parallel - 90° (D) perpendicular and (E) isotropic. Middle row – Images of first recorded (N=1) isochromatic stress fringe observed for each respective disc. Bottom row - Images of the fifth (N=5) isochromatic stress fringe observed for each respective disc.

Visualization of the stress fringes over increasing force loads permitted calibration of the materials for calculation of the stress optic coefficients (*C*). Following the approach of Daniels *et al.* [7], Figure 8 plots the normalized force load with respect the disc diameter (kN/m) against the first six isochromatic stress fringes observed on the discs . The contrast between the sensitivity of the optical properties for the two materials are clearly shown Approximately 150 kN/m force per meter was required to achieve five fringes for the 3D-printed discs. In contrast, close to 500 kN/m was required to achieve the same for the acrylic discs. By using 3D, printing the stress optic coefficient is significantly lowered from its isotropic value. The *C* value was calculated for by using the gradient of the normalized force at wavelength $\lambda$ = 589 nm.

Figure 7 also shows that while the pressure at which the first recorded stress fringe appears depends on the choice of print direction and the relative orientation with respect to the compression plates, the material calibration (Table 2) shows that the mechanical force required remains consistent. [24] VeroClear is an acrylate-based UV curable resin, which deforms like a ductile material [23]. Manufacturer formulations are subject to alteration from time to time, and so a recalibration of the stress-optic coefficient is advised for each new batch.

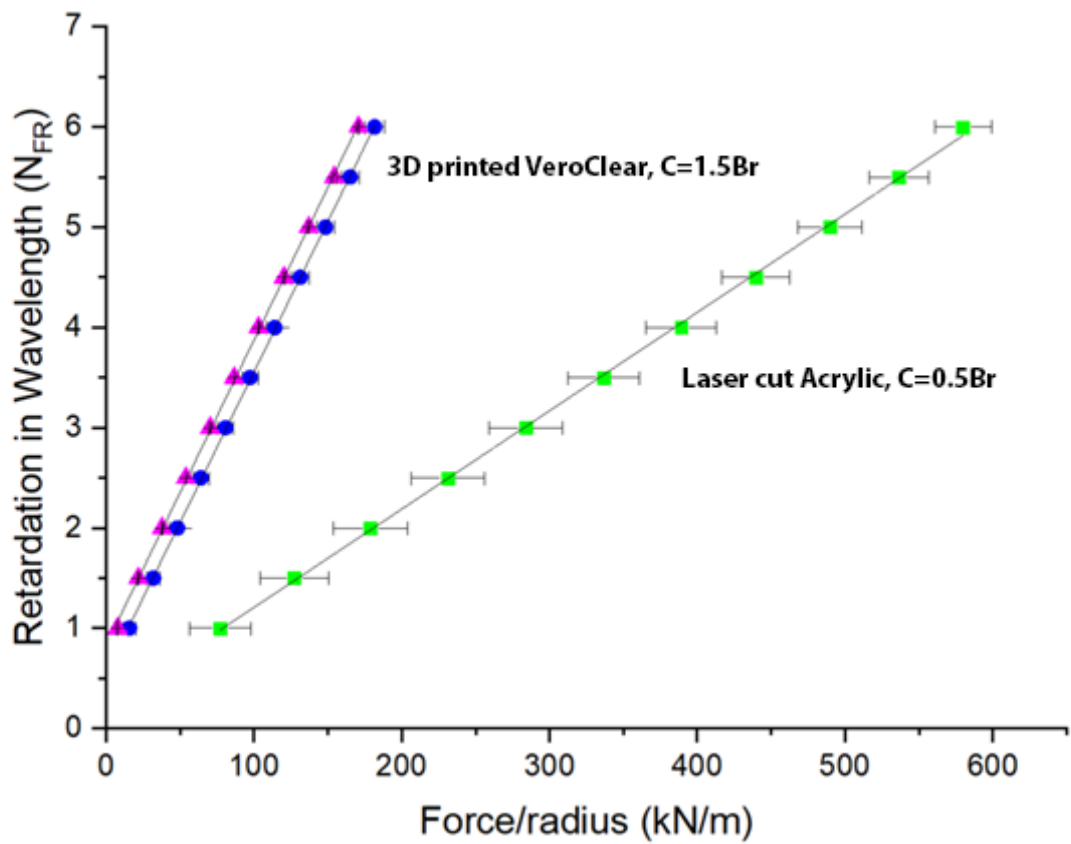

**Figure 8. Graph for the calibration of $F_\alpha$ for the discs in VeroClear and acrylic materials. The 3D printed disc: VeroClear with 'parallel' Δ (average taken for 0°, 45° and 90°) and 'perpendicular' ◯ print layers with respect to the compression plates. The laser cut disc: Acrylic ▫.**

### 3.2 Finite element analysis of the disc stress visualization

We conducted ASTM tensile and compression tests on the VeroClear to determine the mechanical properties of the VeroClear material for the FEA model parameters. Table 1 summarizes the material values generated for VeroClear in this study. We measured no significant difference in the elastic modulus with respect to print layer orientation. Since the optical behavior of the perpendicular printed disc was most in line with the control disc (an isotropic material) the FEA model was treated in the same way.

For the disc shape, the principal stress fields from the stress-optic equation were used to visualize the stress fringes in FEA. It is important to note that FEA produces an iso-surface mesh, which means that the visualized regions of stress apply only to the surface being viewed. Due to the 2D shape, the principal stress remains constant through the cross-section. Figure 9 shows the visual evolution of the FEA disc commencing from a mesh representation, through to the final gradient intensity.

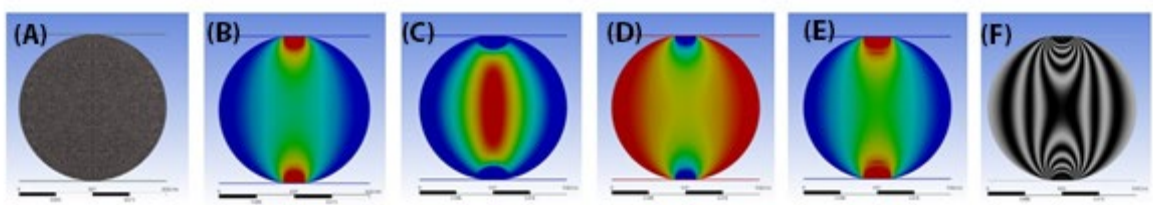

**Figure 9. Evolution of the FEA disc (A) Mesh of the disc (B) Von Mises stress (C) Maximum principal stress (D) Minimum principle stress (E) Difference in principle stress (F) Intensity**

Table 2 compares the experimental and computational data for the 2D disc at $N=5$. The load requirement and displacement needed to obtain $N=5$ in the FEA simulation was in good agreement with the experiment. The slight discrepancy between the force load from the experiment and the FEA could be accounted for the simple linear elastic model used for the material in the FEA, since the compressions done at small displacement. A model which considers more complex material properties and deformation of the disc is likely to produce results closer to the experiments.

Table 1. Material mechanical properties of VeroClear and Acrylic generated according to the ATSM D638 and D695 standards and values reported in literature.

|  | VeroClear | Specimen (Fig. 2) | Elastic Modulus (MPa) | Poisson's ratio |
| --- | --- | --- | --- | --- |
| **ASTM D638 standard tensile test** |  | S1 | 3028 ± 112 | - |
|  |  | S2 | 2788 ± 124 | - |
|  |  | S3 | 2953 ± 102 | - |
| **ASTM D695 standard compression test** |  | S4 | 2586 ± 14 | - |
|  |  | S5 | 2591 ± 21 | - |
| **Simulation data** |  |  | 2500 | 0.38 |
| **VeroClear**[§] |  |  | 2000-3000 [24] | 0.38 [25] |
| **Acrylic**[§] |  |  | 3300 [29] | 0.38 [30] |

[§]Literature values

Table 2. Experimental data for force load measurements taken directly from the Instron for 5 isochromatic fringes (N=5). The compressed discs 3D printed in VeroClear (parallel & perpendicular layers) and for the control disc. Comparative FEA data for VeroClear.

|  | Disc shape (Fig. 6) | Force Load (N) | Displacement (mm) |
| --- | --- | --- | --- |
| **3D-printed material - VeroClear** |  |  |  |
| Parallel | **A, B, C** | 1419 ± 73 | 0.3 ± 0.02 |
| Perpendicular | **D** | 1542 ± 34 | 0.3 ± 0.02 |
| FEA |  | 1666 | 0.2 |
| **Laser cut material - Acrylic** |  |  |  |
| Isotropic | **E** | 4809 ± 106 | 0.8 ± 0.04 |

Figure 10 illustrates the first recorded isochromatic stress fringe (*N=1*) for the 3D-printed disc with parallel orientation 45 degrees to the compression plates, both in the experiment and FEA. Of interest was the asymmetrical fringe pattern exhibited by the 45-degree rotated specimen, as this behavior could not be replicated in the FEA simulations using the linear isotropic material model. Implementing a 2D orthotropic model in FEA and significantly decreasing the elastic modulus in one axis (to mimic the variation in strength depending on the 3D printed layer direction) produced a similar asymmetric pattern. However, this anisotropic material model conflicts with experimental test data, and lacks an accurate shear modulus value. A more detailed investigation into the anisotropic material properties and their effect on photoelastic properties was beyond the scope of this study.

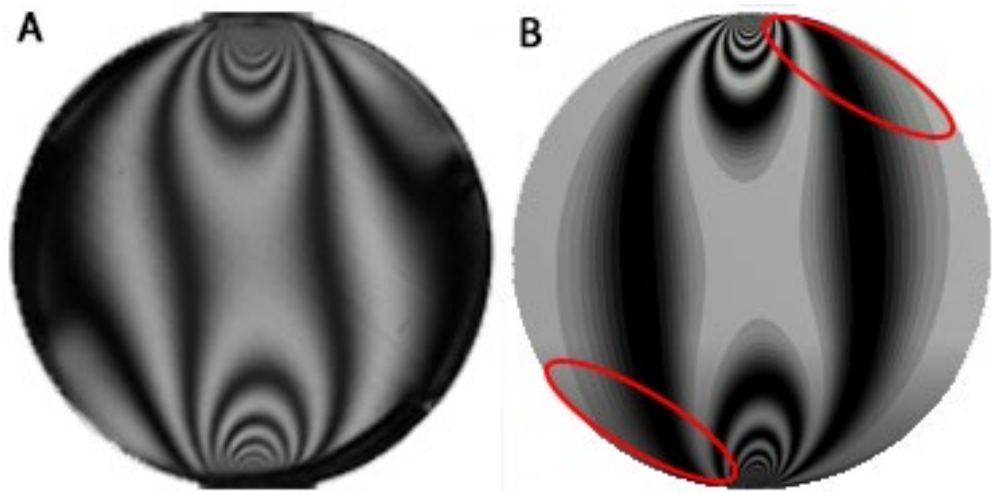

**Figure 10. Image of particle under load to give N=1 for (A) 3D-printed disc with parallel layers at the 45° orientation and (B) the disc in the FEA with the transverse isotropic material model. Parameters used: Orthotropic linear elastic material. Elastic modulus 1: 1500 MPa, Elastic modulus 2: 3000 MPa, Poisson ratio: 0.38, Shear modulus: 800 MPa. The element coordinate system oriented 45° from horizontal.**

3.3     Stress visualization of complex 3D-printed shape

Figure 11 shows the image of (Fig 11A) the real bean, (Fig 11B) the computer-aided model of the bean and (Fig 11C) the final 3D-printed bean in VeroClear. The 3D-printed coffee bean samples were compressed in the upright position (bean line perpendicular to the steel plates).

Compressing the bean on its side (bean line parallel to the steel plates) was possible but the experiments were less repeatable due to instabilities caused by the irregularity of the shape. The coffee bean designs were 3D-printed in the perpendicular orientation (as shown in Fig. 7D) in order to minimize the effect of the anisotropic photoelastic response.

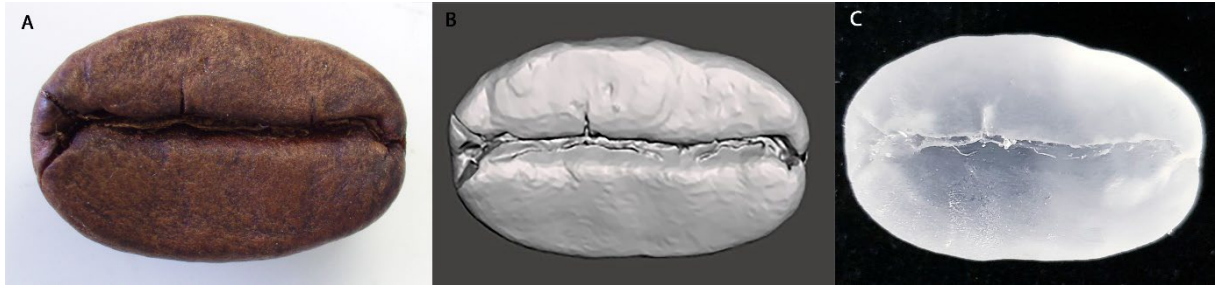

**Figure 11. Image of (A) real coffee bean (B) CAD model of coffee bean and (C) 3D-printed coffee bean. Dimensions of (A) 11 mm length, 4mm thickness and 7 mm height and (B, C) 22 mm length, 14 mm width and 8 mm thickness**.

We observed some dark patches within the photoelastic response of the unloaded 3D printed coffee bean. No obvious stress fringes were observed for the bean particles with voids due to the complexity of the 3D shape throughout the cross-section. However, stress fringes were clearly visualized in a solid particle using existing methods for 2D and pseudo-2D particles. This shadow-like effect could be eliminated by surrounding the particle with media of matching refractive index (*n*). This hypothesis was validated by in silico ray tracing experiments of the uncompressed particle, where a ray was passed through the particle with matching and non-matching *n*, respectively. Media with matching *n* media eliminates the shadow. The shadow is caused by refraction at the multiple faces of the complex shape and not by frozen internal stress, as this effect was not observed in the 3D printed discs. Figure 12 shows ray traced images of the uncompressed coffee bean (with and without internal voids) in a surrounding environment with from matching (left) to non-matching (right) refractive index to the particle; note the existence of a dark patch at the outer edge of both 3D-particles with non-matching refractive index. The shadow on the outer edge of the 3D particle produced by the mis-matched refractive index is accounted to a lensing effect.

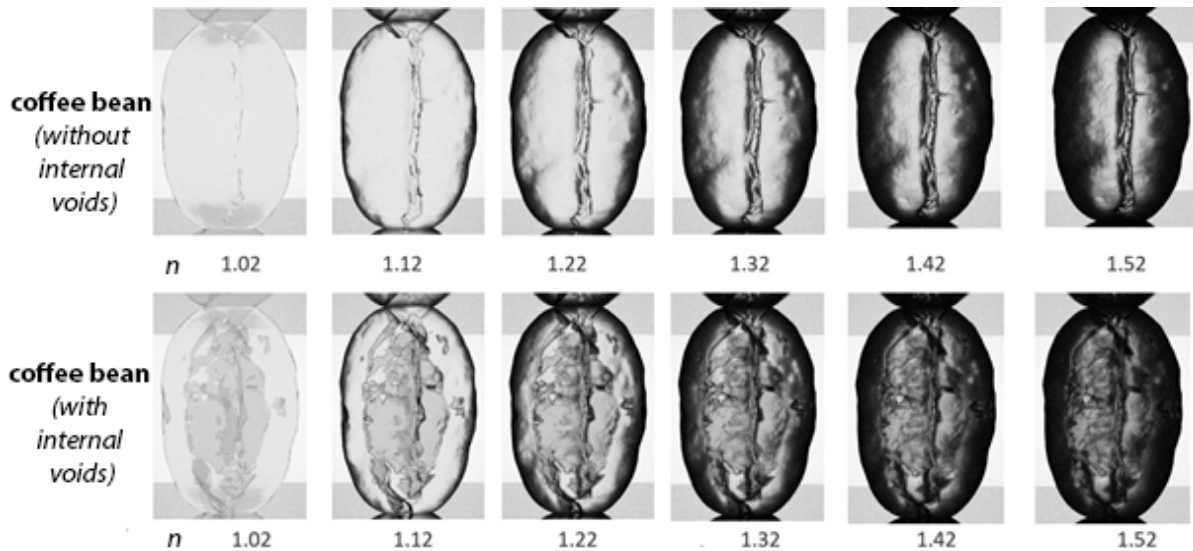

**Figure 12. Ray tracing images of uncompressed coffee bean; Top row – without internal voids, Bottom row – with internal voids; surrounded by (left) media with matching refractive index through to (right) media with non-matching refractive index (same as experiment).**

We performed FEA simulations for the compression of the coffee bean, as shown in Figure 13, with internal voids displaying the Von Mises stress at 0, 1 mm and 2 mm displacement. Furthermore, slices at intervals of 1 mm were extracted to show the stress distribution through the whole bean. The intensity can then be determined at any point in the FEA for comparison to the experimental response. However, the direct comparative experimental setup could not be carried out due to instability of the bean at such fine measurements.

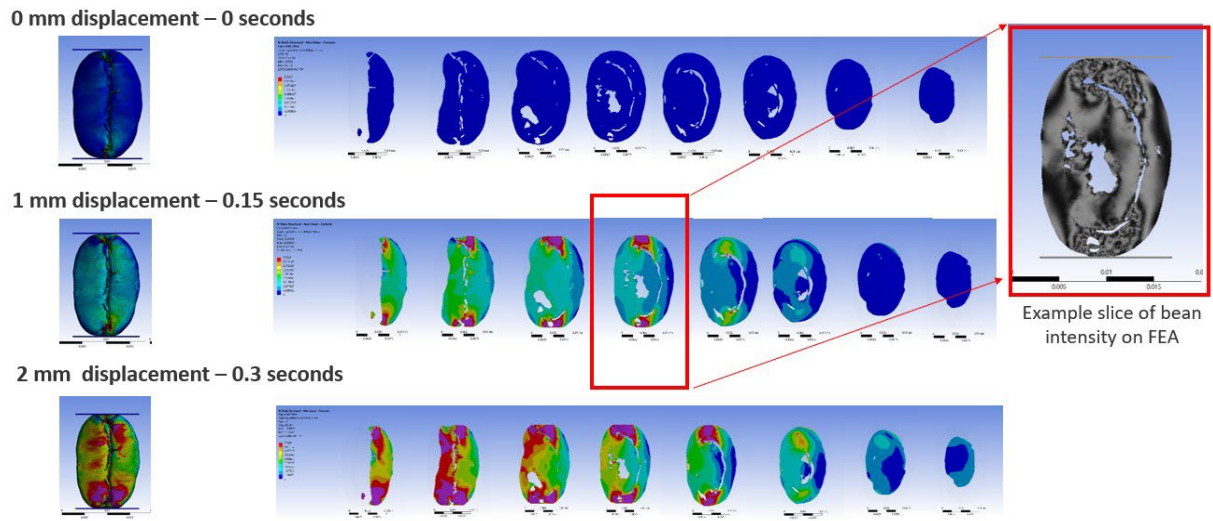

**Figure 13. FEA compression for the coffee bean with internal voids at 0, 1 and 2 mm displacement. Slices taken at every mm throughout the whole bean starting from (left hand side) the front of the bean to (right hand side) the back of the bean.**

In one prior study, an X-ray CT scan of a soil sample with inherent internal fractures was compressed under conditions which captured and locked in the internal stress [31]. Using this technique, slices through the cross-section were able to be taken of the experimental sample after compression and directly compared to the FEA. It is important to note that the sample was a stable shape which could be keep upright without modification to the design or support. Therefore, to try and observe the photoelastic response inside the bean with the internal voids, a planar cross-section of the bean particle was extracted and digitally extruded to 10mm thickness for 3D printing. This was done to aid in understanding the formation of stress fringes around the voids.

For a cross section of the coffee bean with internal voids, the experimental photoelastic response was compared against the FEA intensity. Figure 14A shows the experimental photoelastic compression of the bean slice at a low displacement. The comparative 2-Dimensional FEA analysis of the digitally extruded cross-section demonstrated good agreement with respect to the formation of isochromatic fringes, particularly regions of high stress around the voids (Fig 14B). Additionally, compression of the particle eventually led to particle fracture.

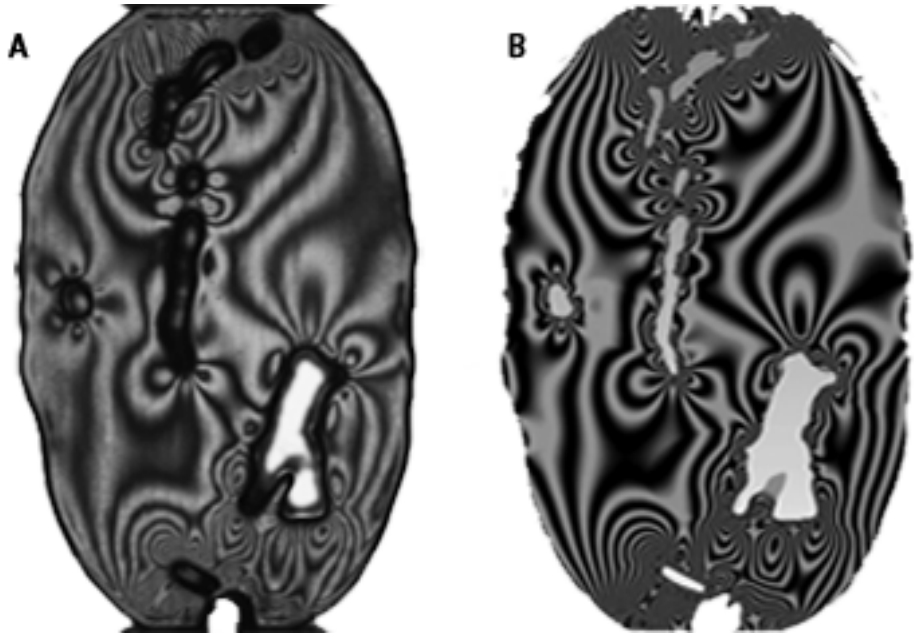

**Figure 14. Photoelastic response of the (A) 3D-printed digitally extruded cross-section of the coffee bean with internal voids and (B) FEA simulation, both to 0.25 mm displacement.**

3.4     Stress quantification of 3D-printed particles using gradient-squared method

There have been a few reports on the stress quantification of complex shaped particles [11, 13, 31] The addition of complex internal voids has not been considered in these methods. Currently there are no solutions to this problem. The gradient squared method ($G^2$) is a method which could be used for such particles as it has proven successful in a variety of circumstances. It is an approach which has been used to measure the empirical 2D stress of particles (mainly discs) by quantifying the sum of the squared gradient of light intensity produced over increasing force loads. Although this method has been used for studying stress birefringence of granular packing in 3 dimensions [11], it has never been applied to 3D-printed particles of complex or porous geometries. The presence of bright voids would need to be excluded from the $G^2$ analysis in order for it to be meaningful; this would be possible with image segmentation.

Semi-quantitative stress measurements have been applied to conventionally flat, but non-circular, particle shapes by Zhao *et al.* [10], and we perform a similar analysis here. Figure 15 presents the plot of the $G^2$ produced from the compression videos of the particles against the respective normalized force loads. Comparing the $G^2$ value for VeroClear and acrylic materials; it is much higher for VeroClear, possibly due to the difference in stress optic coefficient reported above.

The $G^2$ value grows linearly with the applied load (Fig 15D), as expected with this method for photoelastic particles [5, 8]. The $G^2$ value for the 3D-printed 2D disc shapes (Fig 15A & C) exhibited a linear trendline with the applied load above approximately 3 MPa, comparable to the control disc, before reaching a plateau. The initial deviation in the trendline could potentially be due to asperities on the unpolished particle surface, with the particle shifting position during the early stages of compression. As the strain on the 3D-printed discs increased, larger deviations in the $G^2$ value were recorded, which could. be due to effects caused by the material plastically deforming (making the observations not reversible as the load is decreased). This suggests that there is window for meaningful stress interpretation at low loads, before it begins to slightly plateau and ultimately deviate at higher loads.

The $G^2$ value for the 3D coffee bean (Fig 15E) with solid internals does not behave in the same way as the 2D particle shapes. There is an initial deviation in the $G^2$ value before it plateaus, which could be due to a lensing effect caused by the curved structure of the particle, as reported above. However, the 2D extruded cross-section of the coffee bean displays a similar trend to the 3D-printed 2D disc shape. This suggests that the $G^2$ method can be applied to 3D printed particles and is better suited for particles with a constant 2D cross-section, and that the irregularity of the external and internal particle shape is not a limitation for this method. To improve the method [9-11], the fully 3D particles should be observed in a bath containing index-matched solution to avoid lensing effects.

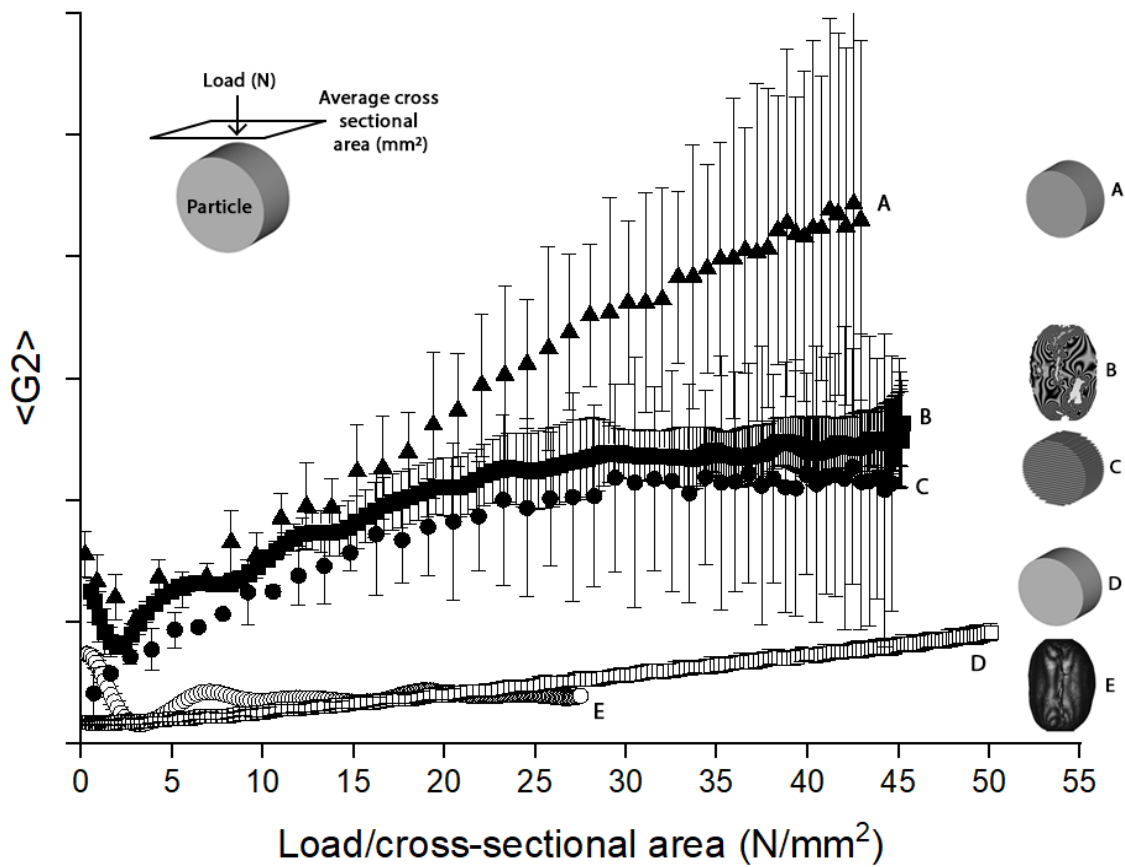

**Figure 15. The gradient-squared ($G^2$) approach used to quantify the amount of light produced for the tested particles. The gradient squared light intensity plotted against the normalized force loads shown for the particles starting from top to bottom of graph: 2D VeroClear disc – perpendicular print layers (A), 2D digitally extruded cross-section of coffee bean (B), 2D VeroClear disc – parallel print layers (C), 2D acrylic disc (D) and 3D solid coffee bean (E).**

## 4      Conclusions

This study explored the feasibility of photoelastic stress measurements within several times of 3D-printed particles. For disc shapes, we found variations in the photoelastic properties due to the relative orientation of the print axis, particle axis, and compression axis. These changes occurred even though the mechanical properties were not significantly affected by the change in orientation of the print layer. In addition, we observed a strong photoelastic response in more complicated 3D shapes, although only a semi-quantitative interpretation was accessible.

We explored the possibility of using a digitally extruded 2D cross-section, and found significant differences compared to the fully 3D bean. We found that, due to mechano-optical artefacts from printing process, the VeroClear material should be recalibrated to the specific print orientation for each specific particle fabrication. FEA results modeled the regions of high and low stress, particularly around the internal voids, with the fringes in good agreement to the experimentally observed fringes. Intricate internal geometries make the stress visualization within 3D shapes difficult to quantify. We also report, for the first time, the $G^2$ method applied to 3D-printed particles of simple and complex geometries for semi-quantification of the internal stress. A linear trend was produced for 2D circular and irregular shapes, although the complexity of 3D shapes, with or without internal voids, requires more sophisticated methods to be developed.

The use of VeroClear as a printing material shows promise, in concert with the development of some additional techniques. Further work to develop its modelling should treat it as an anisotropic material and account for the rotation of polarization due to the material itself, beyond the superimposed photoelastic response. This would allow for more quantitative measurements than are presently possible. With this in place, 3D printing could be used to investigate the photoelastic stress visualization for complex particle geometries, bringing researchers one step closer to being able to understand the behavior of and model the stress within complex geometries.

**Acknowledgements**

This work was supported by the International Fine Particles Research Institute (IFPRI) and an Australian Research Council (ARC) Discovery Project ( DP150100119). KED is grateful for support from the National Science Foundation, grant DMR-2104986. The authors would like to give thanks to Dr Asadul Haque for assistance with X-ray Computed Tomography scans of the coffee bean at Monash University. We also thank Damian Elderfield for his assistance with processing 3D printing files at Deakin University.